\documentclass[a4paper,twocolumn,11pt,accepted=2024-04-17]{quantumarticle}

\pdfoutput=1
\usepackage[utf8]{inputenc}
\usepackage[english]{babel}
\usepackage[T1]{fontenc}
\usepackage{amsmath}
\usepackage{subcaption}
\usepackage{tikz}
\usepackage{hyperref}
\usepackage{makecell}
\usepackage{diagbox}
\usepackage{multirow}
\usepackage{float}
\usepackage{cite}

\begin{document}

\title{Resource-Efficient Real-Time Polarization Compensation for MDI-QKD with Rejected Data}

\author{Olinka Bedroya}
\email{o.bedroya@mail.utoronto.ca}
%\orcid{0000-0003-3758-6909}
\affiliation{Centre for Quantum Information and Quantum Control, Dept. of Physics, University of Toronto, Toronto, Ontario, M5S 1A7, Canada}
\author{Chenyang Li}
%\orcid{0000-0002-5554-1889}
\affiliation{Centre for Quantum Information and Quantum Control, Dept. of Electrical \& Computer Engineering, University of Toronto, Toronto, Ontario, M5S 3G4, Canada}
\affiliation{Department of Physics, University of Hong Kong, Pokfulam, Hong Kong}
\author{Wenyuan Wang}
\affiliation{Department of Physics, University of Hong Kong, Pokfulam, Hong Kong}
%\orcid{0000-0002-8516-9168}
\author{Jianyong Hu}
\affiliation{Centre for Quantum Information and Quantum Control, Dept. of Electrical \& Computer Engineering, University of Toronto, Toronto, Ontario, M5S 3G4, Canada}
\author{Hoi-Kwong Lo}
\affiliation{Centre for Quantum Information and Quantum Control, Dept. of Physics, University of Toronto, Toronto, Ontario, M5S 1A7, Canada}
\affiliation{Centre for Quantum Information and Quantum Control, Dept. of Electrical \& Computer Engineering, University of Toronto, Toronto, Ontario, M5S 3G4, Canada}
\affiliation{Quantum Bridge Technolgies, Inc., (QBT), 100 College St, Toronto, ON M5G 1L5, Canada.}
%\orcid{0000-0002-0340-4989}
\author{Li Qian}
\affiliation{Centre for Quantum Information and Quantum Control, Dept. of Electrical \& Computer Engineering, University of Toronto, Toronto, Ontario, M5S 3G4, Canada}
%\orcid{0000-0003-3550-6581}

\maketitle

\onecolumn
\vspace{-5 mm}
\begin{abstract}
  Measurement-device-independent quantum key distribution (MDI-QKD) closes all the security loopholes in the detection system and is a promising solution for secret key sharing. Polarization encoding is the most common QKD encoding scheme, as it is straightforward to prepare and measure. However, implementing polarization encoding in MDI QKD imposes extra challenges, as polarization alignment must be maintained over both mutually unbiased bases and be maintained for both paths (Alice-Charlie and Bob-Charlie). Polarization alignment is usually done by interrupting the QKD process (reducing overall key generation rates) or using additional classical laser sources multiplexed with quantum channels for polarization alignment. Since low key rates and cost are the two most pressing challenges preventing wide adoption of QKD systems, using additional resources or reducing key rates runs contrary to making QKD commercially viable. Therefore, we propose and implement a novel polarization compensation scheme in the MDI-QKD system that avoids the aforementioned drawbacks by recycling part of discarded detection events. Our scheme evaluates the polarization drift in real-time based on single measurements corresponding to decoy intensities. Our fully automated experimental demonstration maintains the average polarization drift below 0.13 rad over 40 km of spooled fibre (without an insulating jacket) for at least four hours. The average quantum bit error rate is 3.8$\%$, and we achieved an average key rate of $7.45\times 10^{-6}$ bits per pulse.
\end{abstract}
\vspace{5 mm}

\twocolumn
\section{Introduction}

Quantum key distribution (QKD) allows two parties to share a secret key over an insecure channel and is proven to be information-theoretically secure \cite{Security}. However, the devices used in practice may deviate from the idealized models used in security proofs, and this deviation can be exploited to hack the system (see \cite{hacking} for a review).
Measurement-device-independent quantum key distribution (MDI-QKD) \cite{MDIQKD} is a promising solution as it closes all existing and yet-to-be-discovered side-channel loopholes on the detection system, which used to be the most vulnerable part of QKD systems. MDI-QKD has been implemented by many research groups\cite{MDIQKD4, MDIQKD3, MDIQKD1, MDIQKD2, MDIQKD5, MDIQKD6, MDIQKD7, MDIQKD8, MDIQKD9, MDIQKD10,MDIQKD11,MDIQKD12, MDIQKD13, MDIQKD14,MDIQKD15,MDIQKD16,MDIQKD17,MDIQKD18,MDIQKD19,MDIQKD20,MDIQKD21} and has been demonstrated to be feasible and practical. 

In MDI-QKD, the detection is completely handed over to an untrusted third party who is supposed to perform a Bell state measurement on states received from two users. This structure makes MDI-QKD compatible with a multiuser star network topology. In addition, the users do not need a detection system which significantly reduces the setup cost per user.
Furthermore, end-users do not have to be directly connected; they only need a connection to a central detection node to share a secret key. These advantages make the idea of the MDI-QKD network very favourable in practical applications. However, one of the significant challenges in realizing an MDI-QKD network is maintaining the indistinguishability of the signals sent by different users required for successful Bell state measurement. So far, there has been only one MDI-QKD network implementation which used time-bin phase-encoding \cite{MDIQKD5}.

MDI-QKD can be implemented using polarization, time, or phase encoding. An MDI-QKD network with polarization encoding will have two advantages: (1) State preparation and measurement are simple. Unlike phase and time-bin encoding, polarization encoding does not involve the complications of stabilizing interferometers. For time-bin schemes, the interferometers should have matching arms for each user, and when multiple users are involved, the interferometry alignment for many users becomes impractical. Therefore, multiuser implementation using polarization encoding is more feasible. (2) Polarization encoding is highly favoured for free space QKD. Using polarization encoding facilitates a heterogeneous QKD network where free-space and fibre channels can co-exist. 

However, polarization encoding in fibre-based implementations comes with its challenges. Firstly, polarization variations due to the temperature fluctuations and birefringence of the fibre are inevitable and affect QKD performance \cite{Drift}. Secondly, implementations using polarization encoding are more demanding as the indistinguishability condition of Bell state measurement requires a nearly perfect alignment of both rectilinear and diagonal bases of two users to Charlie's bases. In contrast, polarization alignment is still necessary for other schemes to achieve high interference visibility, but only for each user's principal polarization. Therefore polarization compensation is the first step toward realizing an MDI-QKD network.

In available QKD literature, polarization drift compensation over the link always comes at a significant cost. The drawbacks fall into one of three categories: (1) The key sharing is interrupted. These interruptions are either invoked periodically \cite{PolComp4,PolComp5,PolComp6,PolComp9,PolComp24,PolComp13}, or triggered by an increase in QBER \cite{PolComp12,PolComp13}.  (2) Additional resources are required to multiplex reference polarization pulses with the signal. The reference pulses are multiplexed in either the wavelength domain \cite{PolComp23,PolComp22,PolComp19,PolComp10,PolComp18}, or in the time domain \cite{PolComp21,PolComp11,PolComp3,PolComp18}. (3) A fraction of quantum pulses are sacrificed for tomography\cite{PolComp16,PolComp20,PolComp25}.   
We are aware of only one exception to the abovementioned schemes where real-time compensation was implemented based on the data revealed in the privacy amplification \cite{PolComp17}. Unfortunately, this scheme does not apply to MDI-QKD as privacy amplification is based on coincidences, and the polarization drift of individual users cannot be inferred from it. To date, the MDI-QKD experiments have either incorporated a compensation scheme mentioned above or implemented a passive stabilization scheme where fibres are environmentally isolated, and manual compensation is used when the polarization drifts become large. There have been six MDI-QKD experiments with polarization encoding \cite{MDIQKD3,MDIQKD4,MDIQKD11,MDIQKD15,MDIQKD19,MDIQKD20}, among which only \cite{MDIQKD11,MDIQKD19} have automated polarization compensation schemes. Among the time-bin or phase encoding MDI-QKD experiments with active compensation, either realignment is invoked periodically \cite{MDIQKD1,MDIQKD9,MDIQKD10,MDIQKD12,MDIQKD21} or additional polarization beam splitter and/or single-photon detectors are used \cite{MDIQKD2,MDIQKD5,MDIQKD6,MDIQKD7,MDIQKD14,MDIQKD17}.

The three categories of compensation schemes mentioned above either require extra equipment or reduce the key sharing cycle. In an ideal MDI-QKD network, it is beneficial to keep the nodes as simple as possible so that adding a user could be feasible in terms of cost and practicality. Moreover, since the users share the detection system, it would be ideal not to reduce the key sharing cycle for compensation. 
We propose and implement a polarization compensation scheme that avoids all the abovementioned drawbacks using some of the MDI-QKD's discarded detections. Recycling discarded counts in QKD is not a new idea \cite{discarded}, and here we do it for polarization stabilization. This compensation can be done in real-time without interruption, extra equipment, or sacrificing any quantum signal intended for key sharing for compensation and allows us to retain a simple structure for a user node without adding additional equipment for the compensation. Therefore, we believe that our scheme is favourable for an MDI-QKD network. 

In this paper, our scheme is specifically discussed for polarization encoding MDI-QKD which is the most demanding in polarization alignment among the MDI-QKDs. However, the idea presented here of recycling decoy detections can be applied to other QKD schemes and protocols \footnote{Please refer to the “Future work” section in \cite{thesis}}.

\section{Method}

Our polarization compensation scheme solely relies on the discarded data of the QKD protocol. The security of MDI-QKD is based on the Bell state measurement, and specific coincidence detections mark a successful Bell state projection \cite{MDIQKD}. These coincidences are collected for key distribution, and all single events are discarded. Additionally, detection events associated with decoy intensities \cite{decoy} do not contribute to the key generation. Table \ref{tab1} summarizes the combination of transmitted weak coherent states that, if not empty, should result in a successful Bell state measurement and key generation. In decoy-state QKD, the vacuum state is usually used as one of the decoy states as it allows the users to estimate the background rate. When one user sends a vacuum decoy state, the detection result provides direct information about the polarization state prepared and sent by the other user. The main idea of our scheme is to use the single measurements corresponding to these transmitted states to actively evaluate the polarization drift based on singles' error rates and run compensation independently for each use. Such instances are listed in table \ref{subtab2}. As shown in table \ref{subtab2}, given the optimized decoy probabilities for our experiment, 25.5$\%$ of transmitted states that were previously discarded could be recycled and used to estimate the polarization misalignment of the user's bases to Charlie's bases. To put this number in perspective, only 3.4$\%$ of transmitted states could contribute to the secret key (table \ref{subtab1}). Additionally, if we look at all the single photon detections in our experiment,  10.6 $\%$ could be recycled for each user's polarization alignment estimation, and in total, 21.2 $\%$ of detections that were otherwise discarded could be recycled for polarization compensation. The compensation can be done in real-time without any interruption, extra equipment, or sacrificing any quantum signal intended for key sharing for the sake of compensation.

We briefly summarize the step-by-step procedure to see where our compensation scheme fits within a round of polarization-based MDI-QKD.
\begin{enumerate}
\item State preparation - Alice and Bob select a random bit, basis and intensity (signal or decoy) for each of their pulses and prepare their phase-randomized weak coherent pulse accordingly. 
\item They send their states to Charlie, who performs Bell state measurements on the arriving pulses and publicly announces the results and the measurement basis.
\item Key sifting - Alice and Bob publicly announce their bases and intensity settings. The sifted key is generated from the secret bits used for transmissions where both users used the same polarization basis and the signal intensity, and the transmitted states resulted in a successful Bell state measurement. The decoy data corresponding to successful measurements where both users had the same basis are used for channel estimation.   
\item Polarization alignment information - In addition to the shared public information in the previous step, the users share the polarization bit corresponding to failed measurements for the compensation scheme. Charlie recycles information from some of the failed Bell state measurements to estimate the polarization misalignment of every user, as explained in the rest of this section. Then, he announces the users' polarization misalignments so that they can realign their polarization in real-time. 
\item Users perform error correction and privacy amplification on their sifted key.
\end{enumerate}

\begin{table*}[htb]
            \begin{subtable}{0.4\linewidth}
            \centering 
			\begin{tabular}{|c c| c|}   
			\hline 
			Alice & Bob  \\ [1ex]     
			\hline    
			$|H_\mu\rangle$ & $|V_\mu\rangle$ \\  [0.6ex]
			$|V_\mu\rangle$ & $|H_\mu\rangle$  \\  [0.6ex]
			\hline    
			\multicolumn{2}{|c|}{\small{Total Probability} $\frac{1}{8}P_{\mu}^2$} \\  
			\hline
			\end{tabular}%                  
			\caption{All usable combinations for key generation when Alice and Bob's signals are indistinguishable, and their bases are perfectly aligned to those of Charlie. When using weak coherent pulses, only rectilinear basis is used for key generation \cite{MDIQKD}.}\label{subtab1}
           \end{subtable}%}
           \hspace*{1.5em}
           \begin{subtable}{0.5\linewidth}
			\centering      
			\begin{tabular}{|c c|c|}   
			\hline 
			Alice & Bob  \\ [0.6ex]     
			\hline    
			$|H_\mu\rangle or |H_\nu\rangle$&$|H_\omega\rangle, |V_\omega\rangle, |D_\omega\rangle or |A_\omega\rangle$ \\  [0.6ex]
			$|V_\mu\rangle or |V_\nu\rangle$&$|H_\omega\rangle, |V_\omega\rangle, |D_\omega\rangle or |A_\omega\rangle$ \\  [0.6ex]
			$|D_\mu\rangle or |D_\nu\rangle$&$|H_\omega\rangle, |V_\omega\rangle, |D_\omega\rangle or |A_\omega\rangle$ \\  [0.6ex]
			$|A_\mu\rangle or |A_\nu\rangle$&$|H_\omega\rangle, |V_\omega\rangle, |D_\omega\rangle or |A_\omega\rangle$ \\  [0.6ex]
			\hline    
            \multicolumn{2}{|c|}{\small{Total Probability} $P_{\omega}-P_{\omega}^2$} \\  
			\hline
			\end{tabular}%
		\caption{Combinations of transmitted states that could reveal Alice's polarization alignment through single detections. No key is extracted from these combinations, and the results are otherwise discarded.} \label{subtab2}
           \end{subtable}%
        \caption{Overview of measurements in MDI-QKD. Here, we assume two decoy-state implementations, with signal intensity $\mu$ and decoy intensities $\nu$ and $\omega$, where $\omega$ is close to vacuum. For optimized probabilities of transmitting each of these intensities $P_\mu=0.52, P_\nu=0.33$ and $P_\omega=0.15$, 12.8$\%$ of states could be used similarly to infer each user's polarization alignment (b). These add up to 25.5$\%$ of transmitted states, which were previously discarded, but could be recycled and used. In comparison, only 3.4$\%$ could contribute to the secret key (a).}\label{tab1}
       \end{table*}%
        
\begin{figure}[t]
  \centering
  \includegraphics[width=\columnwidth]{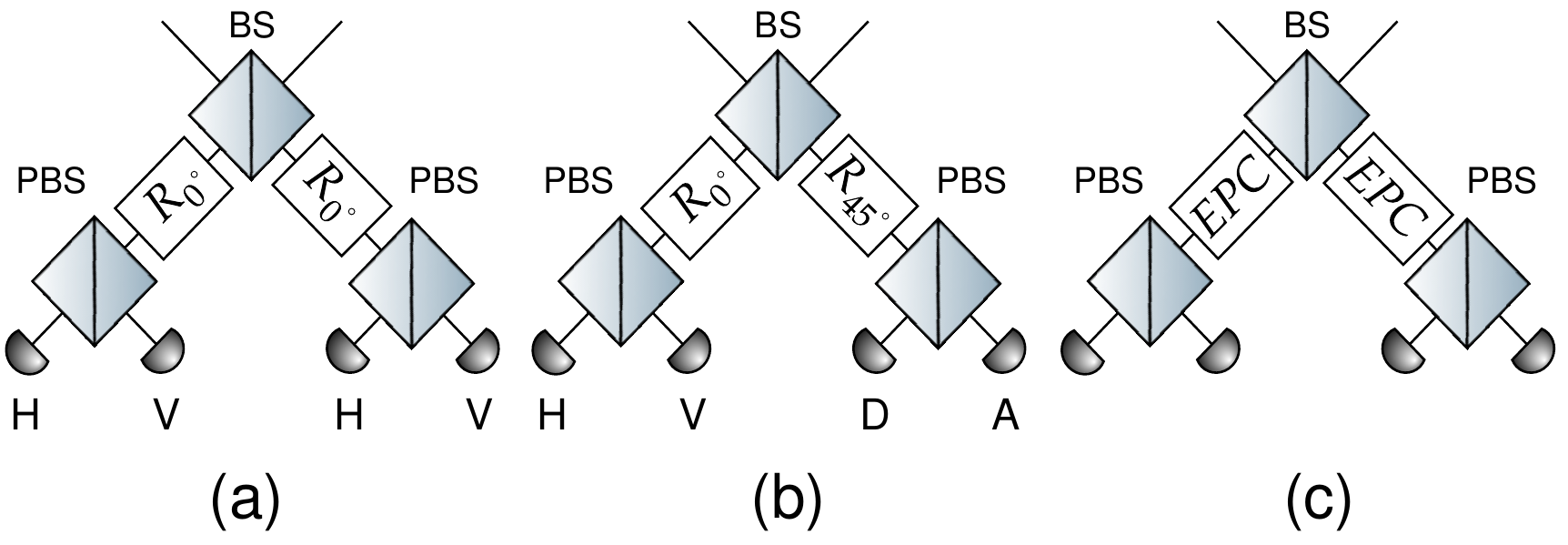}
  \caption{(a) The measurement setup in the original MDI-QKD scheme where both polarization beam splitters (PBS) are aligned to the rectilinear basis. (b) Both bases are defined at the measurement node by setting one of the PBS to the diagonal basis; One of the distinguishable Bell states has to be dropped to maintain the QBER due to multiphoton weak coherent pulses low. (c) Both bases are defined at the central node as the axes of both PBSs are rotated simultaneously and periodically between rectilinear and diagonal bases using electronic polarization controllers (EPCs). The active switching might be more complicated than a passive modification in (b); however, it comes at no cost to the rate of successful Bell projections.}
  \label{fig:Charlie}
\end{figure}

\quad To simultaneously track and maintain the alignment of each user's rectilinear and diagonal bases, we have proposed a slight modification to the measurement node. In the original MDI-QKD paper, the Bell-state measurement implementation includes two polarization beam splitters projecting onto the rectilinear basis. This design defines the rectilinear reference basis at the measurement node, and each user can independently align to it (figure \ref{fig:Charlie}a). However, this is insufficient for polarization encoding MDI-QKD, and we also need the diagonal bases of users to be aligned. To our knowledge, there has not been a formulated solution for this challenge without using additional resources (reference pulses or tomography equipment). Traditionally at least two polarization controllers, one before each polarization beam splitter, should be used to align the optical setup for Bell state measurement (Figure \ref{fig:Charlie}a). A slight modification to the measurement setup using these two polarization controllers can address this challenge. There is no need to add additional controllers to the experimental setup. One approach is to rotate the axis of one of the polarization beam splitters in the measurement setup. As shown in Figure \ref{fig:Charlie}b, this modification defines both rectilinear and diagonal bases at the measurement node. This solution is simple yet has a drawback when using weak coherent pulses. One must forgo one of the Bell states to keep the QBER due to multiphoton pulses low (reducing the key rate by half). Our solution is the simultaneous rotation of the axes of both polarization beam splitters (Figure \ref{fig:Charlie}c), which does not reduce the key rate. The switching frequency (period $\tau$) can be optimized for different setups; it should be low enough to allow collection of sufficient data to estimate the drift and sufficiently high to keep track of polarization drift in real-time. In our experiment, the switching is done every 15 seconds. This modification to the measurement setup allows us to infer the polarization drift in both bases using the aforementioned single detections. 

In the compensation scheme, our goal is to retain the initial polarization alignment. We assume each user can create four perfect polarization states (qubit assumption). Any unitary polarization transformation in the fibre affects the alignment of both linear bases. Since we only detect the projections on two bases and do not perform a full tomography on the transmitted states, we do not have complete information about the unitary transformation. However, obtaining complete information is unnecessary in our fibre-based implementation. Because low-loss fibre-based polarization controllers (including the EPCs used here) do not provide information on the polarization rotation they perform, one cannot use these PCs to perform a predetermined polarization rotation. Instead, one can only use them to minimize or maximize a particular measurement reading. Minimizing the misalignments in both linear bases is equivalent to the total compensation of an arbitrary linear transformation. 
Given the above constraints, we took the approach to minimize the polarization misalignments for both the Z and X basis ($\theta_Z$ and $\theta_X$) by minimizing their average, for which there exists only one global minimum where both angles are zero. Our algorithm iteratively adjusts different fibre squeezers for this minimization. Four squeezers provide adequate degrees of freedom for simultaneous compensation of both rectilinear and diagonal bases.

\begin{figure}[t]
  \centering
  \includegraphics[width=\columnwidth]{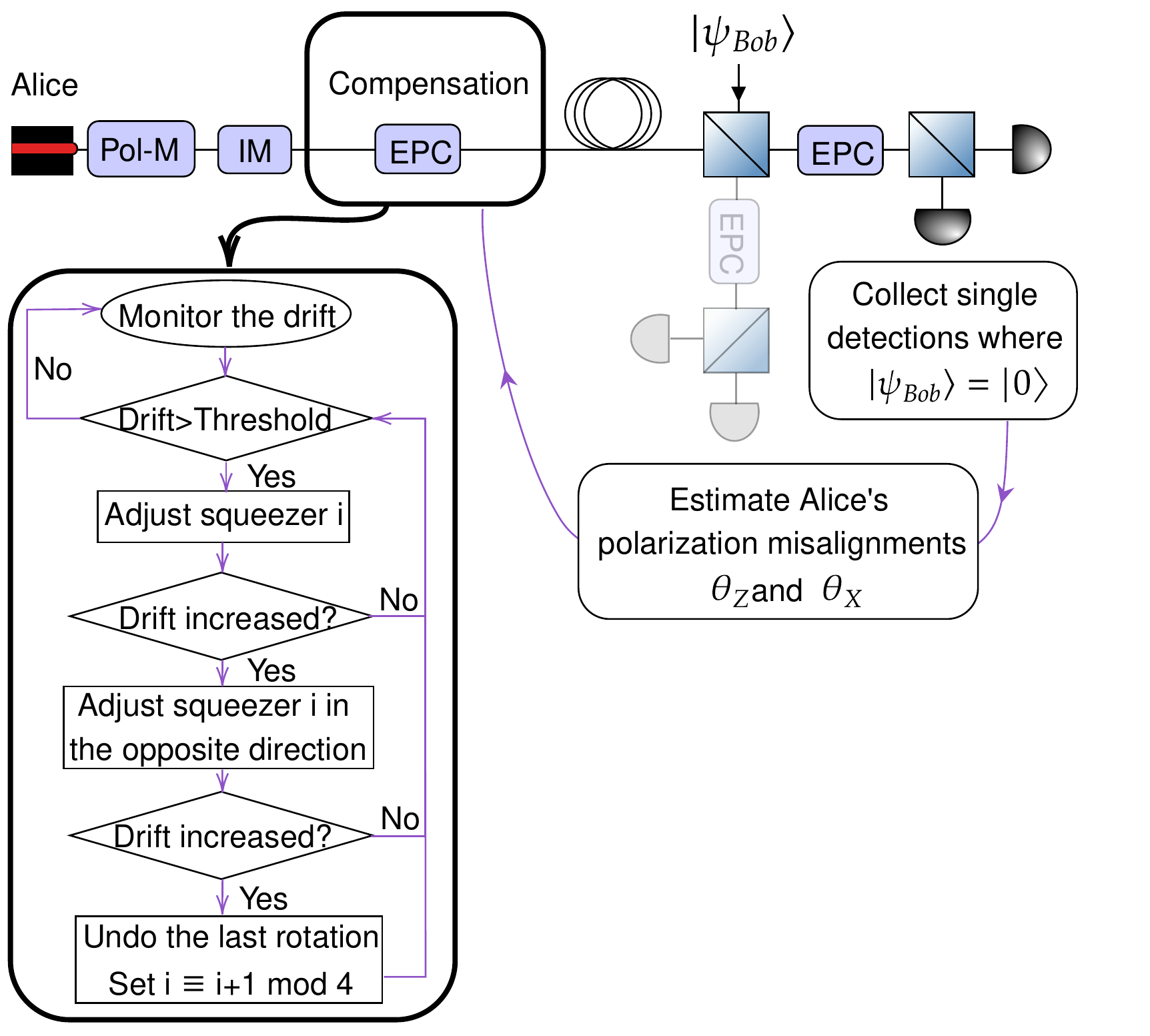}
  \caption{The polarization compensation process. Each user's EPC has four fibre squeezers. The compensation programme keeps adjusting the four squeezers in a cyclic order until the drift is below the threshold.}
  \label{fig:FlowChart}
\end{figure}

Let us describe our compensation scheme as a control loop as shown in Figure \ref{fig:FlowChart}. In that case, the system is the fibre spool between the user and measurement node, the feedback measurements are the two polarization misalignments $\theta_Z$ and $\theta_X$ estimated based on single detections, the input error signal to the controller is their average with the desired value zero, and the control actuator is an electronic polarization controller with four fibre squeezers. We rely on single detections to estimate the misalignment angles $\theta_Z$ and $\theta_X$. Ordinarily, after Charlie announces the successful Bell state measurements in MDI-QKD, Alice and Bob announce their bases and intensity settings for sifting and post-processing. Furthermore, we require Charlie not to discard the single detections right away and announce the single detections when one user has transmitted a vacuum. Then, we request Alice and Bob to reveal their transmitted polarizations corresponding to those timestamps so that Charlie can calculate and announce the polarization misalignments. Note that this revealed information does not correspond to coincidences and does not compromise the key's secrecy. These are the data specified in table \ref{subtab2}. Analyzing the projection outcome of each row provides information about how that specific polarization state of Alice has rotated. Our modified measurement setup allows us to record the projections in both rectilinear and diagonal bases. We estimate the misalignment of each polarization state based on the single detections error rate over a collection time $t_{\text{collection}}$ as $\theta \approx \arcsin(\sqrt{N_{\text{erroneous}}/N_{\text{max}}})$; here $N_{\text{max}}$ is the maximum expected count rate which is intensity-dependent. The compensation will launch if the misalignment in either of the bases exceeds a preset threshold.

\section{Experiment}

\begin{figure}[t]
  \centering
  \includegraphics[width=\columnwidth]{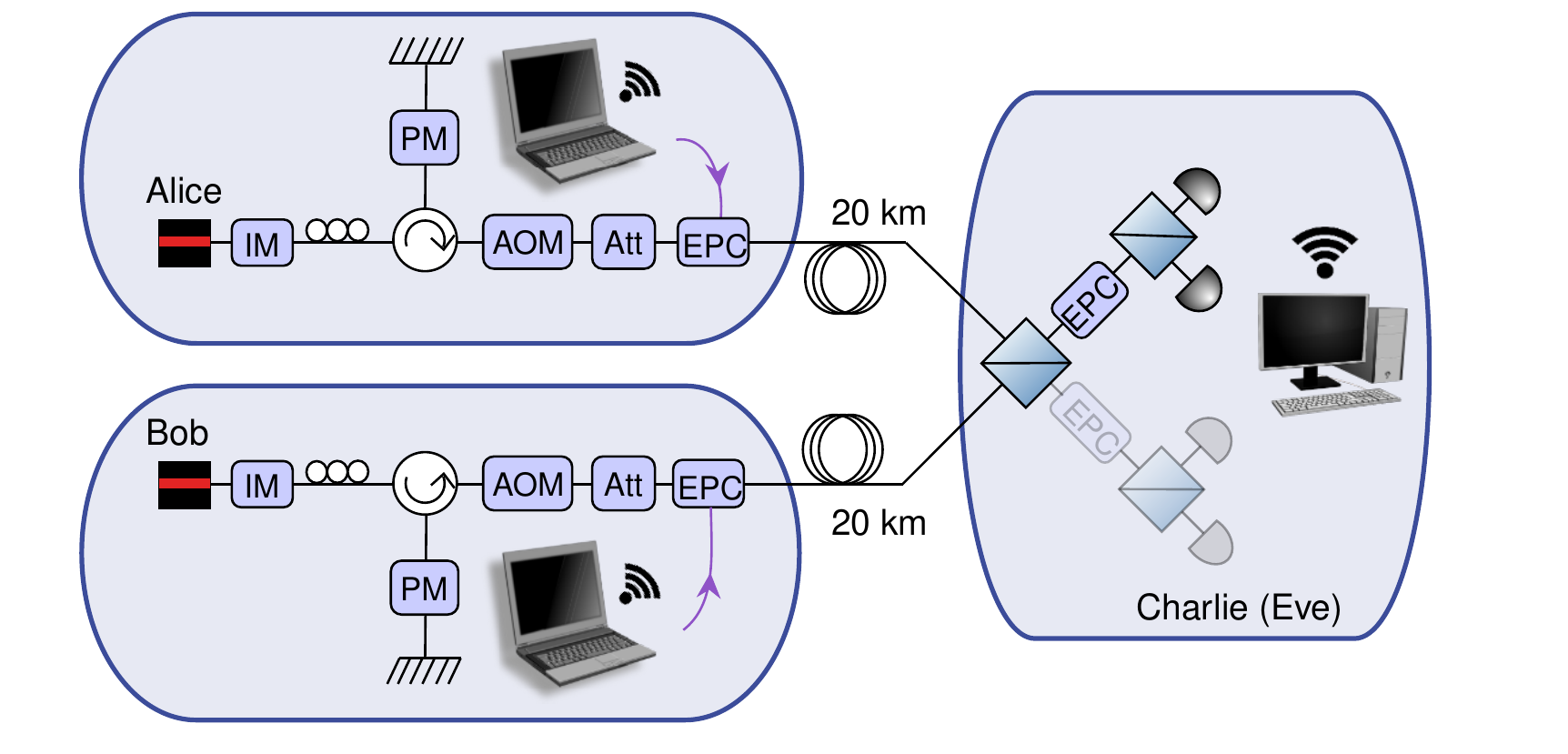}
  \caption{Experimental setup: Attenuator (Att), Intensity Modulator (IM), Phase Modulator (PM), Electronic Polarization Controller (EPC), Acousto-Optic Modulator (AOM). Charlie announces the estimated polarization misalignment of each user over TCP/IP connection, and the users locally run the compensation.}
  \label{fig:experiment}
\end{figure}

The experimental setup is shown in Figure \ref{fig:experiment}. Each user (Alice and Bob) possesses a CW laser (Clarity-NLL-1542) independently locked to a molecular absorption line at 1542 nm such that the frequency difference between the two sources is below 10 MHz. Using an intensity modulator (IM), the users generate coherent pulses with a full width at half maximum of 1.5 ns at a repetition rate of 10MHz. Polarization modulation of the pulses is done using a phase modulator based on the structure proposed in \cite{PolM} to mitigate the effect of thermal fluctuations. We use acousto-optic modulators to adjust the intensity ratio of the optical pulses to generate decoy states. A variable optical attenuator is used to attenuate the light at the output of the users' setup. Each user uses an electric polarization controller (EPC - General Photonics PolaRITE III) with four fibre squeezers for initial alignment at the beginning of the key sharing session and active compensation throughout. These controllers do not impact the prepared BB84 polarization states' quality but are used to maintain their alignment throughout the channel for successful measurement. 

\begin{figure}[H]
  \centering
  \includegraphics[width=0.65\columnwidth]{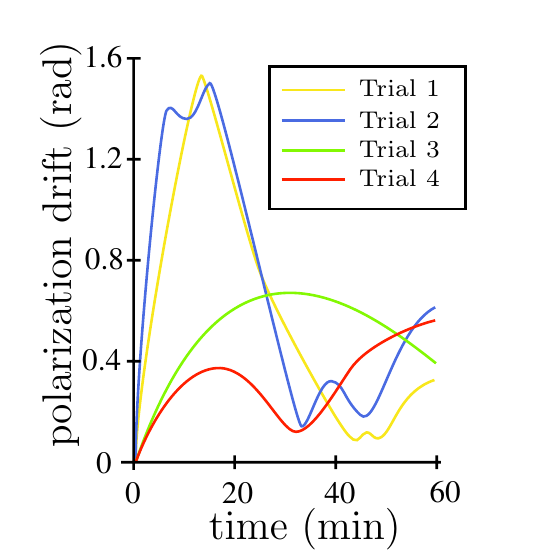}
  \caption{Monitoring polarization alignment over 40 km of unisolated fibre without polarization compensation. Significant Polarization drifts were observed, which call for active compensation in fibre-based QKD systems.}
  \label{track1}
\end{figure}

Each user is connected to the central measurement node via a 20 km fibre spool. The fibre has a 250-um standard acrylate buffer layer and is exposed to the lab environment without thermal or vibrational insulation. Besides the lack of protection by a jacket, since the spooling radius of curvature is small compared to fibres laid out in the field, the spooled fibre in a lab environment can suffer more polarization instability due to environmental disturbances than field fibre \cite{coiled,coiled2}. Sample measurements of polarization drift without compensation in our fibre spools over time are presented in Figure \ref{track1}. At this center node, a third user, "Charlie," performs Bell state measurement, which demands indistinguishability of states received from Alice and Bob. The sources can be deemed spectrally indistinguishable given their pulse width (GHz bandwidth) and the relatively minor frequency difference between the two lasers (10 MHz). Temporal indistinguishability is achieved by adjusting the arrival time of the pulses from two users with 15 ps resolution. Because of the limited availability of detectors, we used only two detectors in Charlie's Bell state measurement setup in a configuration shown in Figure \ref{fig:experiment}, and hence we can only distinguish one Bell state $|\Psi^+\rangle=(|H\rangle|V\rangle+|V\rangle|H\rangle)/\sqrt(2)$ \cite{Bell}. The measurement setup consists of a beam splitter, a polarization beam splitter, an EPC to switch measurement bases, two free-running InGaAs/InP single-photon detectors (ID220), and a time interval analyzer module (ID900) to record the timestamps of the detections. 

Our entire setup is assembled in one laboratory. However, three different computers are used for Alice, Bob, and Charlie to mimic a practical implementation. The three computers use TCP/IP communication over WiFi. Charlie estimates each user's misalignment based on the average single detection error rate (corresponding to Table \ref{subtab2}) over a measurement period $t_{\text{collection}}$. Next, Charlie announces the estimated misalignment to each user. If misalignment in either of the bases exceeds a preset threshold, the automatic compensation on users' EPCs is triggered locally. Each user's EPC (General Photonics PolaRITE III) has four fibre squeezers and is computer-controlled. The compensation program keeps adjusting the four squeezers in a cyclic order parallel to the key sharing until the drift is below the threshold. This program automates how an experimentalist would have intuitively used a manual polarization controller with multiple control knobs.

Three tunable parameters influence the performance of the compensation scheme and can be adjusted based on the experimental setup: the data collection time, the polarization rotation ratio, and the threshold to start the compensation. Below, we explain how we set these parameters:\\

\textbf{Collection time}: There is a trade-off between the response time for real-time drift mitigation and the precision of drift estimation. To compensate for the polarization drift accurately, we need to measure the polarization misalignment precisely based on single detections error rate ($\theta \approx  \arcsin(\sqrt{N_{err}/N})$). We collect data over time intervals to estimate the drift during that interval. Any random polarization fluctuation during the collection time would be averaged out, while consistent polarization drift in one direction would still be captured. However, even if the drift remained exactly constant, the finiteness of the measurement imposes a fundamental limit on how well we can estimate the drift. We can reduce this uncertainty in the drift estimation by increasing the number of detections, i.e. increasing the measurement time. 

We can use Chernoff bound to quantify the number of detections needed to estimate the polarization drift with any desired precision\footnote{We use random sampling theory to estimate the mean of the polarization drift. In principle, Eve could affect the estimated mean by denial-of-service attacks. However, such attacks do not affect the security of QKD, and even in such cases, the Chernoff’s bound still applies \cite{chernoff2}.}. We can model every detection with a coin flip with binary outcomes 0 and 1 corresponding to whether the polarization measurement result matched the transmitted polarization or not. Given a coin bias $p$, these two outcomes happen with probabilities $p$ and $1-p$. For a fixed $N$, we can view the number of erroneous detections $N_{err}$ as a random variable with mean value $Np$ where $p$ is the coin bias $p=\sin(\theta)^2$. We can quantify the quality of our estimation of the actual bias $\hat{p}$ with two parameters: (1) the relative error  $\varepsilon=|p-\hat{p}|/\hat{p}$ and (2) an upper bound ($1-\delta$) on $\Pr[|p-\hat{p}|\leq\varepsilon p]$ which is known as confidence. Using two-sided Chernoff bound\footnote{Multiplicative Chernoff bound\cite{Chernoff} implies $\Pr[|X-\mu|>\epsilon\mu]<(e^\epsilon/(1+\epsilon)^{1+\epsilon})^\mu+(e^{-\epsilon}/(1-\epsilon)^{1-\epsilon})^\mu$ which is always smaller than $2\exp(-\frac{\epsilon^2\mu}{2+\epsilon})$.} we can write $\Pr[|p-\hat{p}|\geq \varepsilon p]\leq 2 \exp(-N\hat{p}\varepsilon^2/(2+\varepsilon))$. Using this inequality, we can see that, for example, if we want to estimate polarization misalignment with a precision of 0.03 rad\footnote{The compensation scheme is activated when estimated polarization misalignment angle is larger than $\theta_{treshold}=0.1$ rad, and the improvement in every step can be an order of magnitude smaller if the axis of the squeezer is not optimally aligned. The precision of 0.03 rad ensures such improvements are detectable.} ($\hat{p}=0.0009$), within an error of $\varepsilon=1/2$, and a 70$\%$ confidence ($\delta=0.3$) we would need at least 21.1 k detections; this is equivalent to 15 seconds of collection time in our experiment since the data rate available to estimate the drift is $1.4\times 10^{3}$ counts per second per basis. After many rounds of trial and error, we found that a collection time of 10-15 seconds seemed to work best for our experiment and proved to be most successful and versatile in maintaining the polarization with longevity. The results in the paper correspond to a collection time of 15s. 

\textbf{Polarization rotation ratio}: The voltage applied to the EPC changes the birefringence of fibre and rotates the polarization state. We adjust the voltage to achieve a rotation proportional to the misalignment angle measured in every compensation step. This proportionality constant $\alpha$ is experimentally tunable. A small $\alpha$ might render the compensation slow and ineffective. On the other hand, a large $\alpha$ might result in an overshoot of QBER if the polarization is rotated in the wrong direction.\\
One might naively expect that $\alpha=1$ provides the optimal compensation. However, even $\alpha=1$ could be too large. This is because, depending on the waveplate axes of the EPC, it could be that any rotation performed by the EPC changes the polarization on the Poincare sphere in a direction perpendicular to the desired orbit. These points suggest that there should be an optimal $\mathcal{O}(1)$ range for the parameter $\alpha$.\\
Assuming the polarization drift in each step is random and uncorrelated with the EPC's waveplate's axes, the numerical simulation shows that $[0.4-0.7]$ is an optimal range for $\alpha$. Experimentally, we observed that $\alpha\in[0.4-0.55]$ works best for compensation with longevity. For the reported results, a coefficient of 0.55 is used for compensation.

\textbf{Threshold}: If the measured drift exceeds a certain threshold value, the compensation process will automatically be triggered. The optimal threshold depends on three factors: the average polarization drift rate (less than 0.003 rad/s for our unisolated 20 km spool), the standard deviation of estimated misalignment angles due to the limited time of measurement, and the precision of the compensation actuator. The threshold should be low to minimize the effect of polarization misalignment on QBER and yet high enough to distinguish compensation performance from the uncertainties associated with the three factors mentioned above. We have found that a threshold $\mathcal{O}(0.1 rad)$ (which is an order of magnitude greater than experimentally measured fluctuations $\mathcal{O}(0.01 rad)$) works with our experimental setup. The threshold for the experimental results reported here was set to 0.13 rad.

\section{Results}

\begin{figure*}[h]
\begin{subfigure}{2\columnwidth}
  \centering
  \includegraphics[scale=0.5]{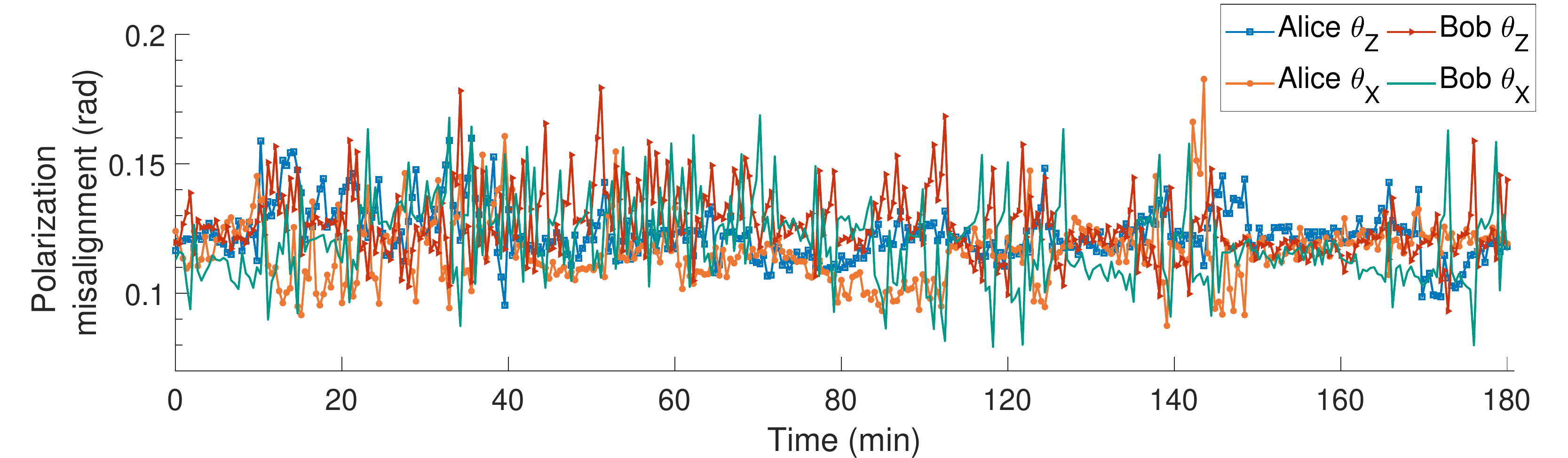}
  \captionsetup{width=9.5cm}
  \caption{\quad}
  \label{track2}
\end{subfigure}%

\begin{subfigure}{2\columnwidth}
\centering
\includegraphics[scale=0.5]{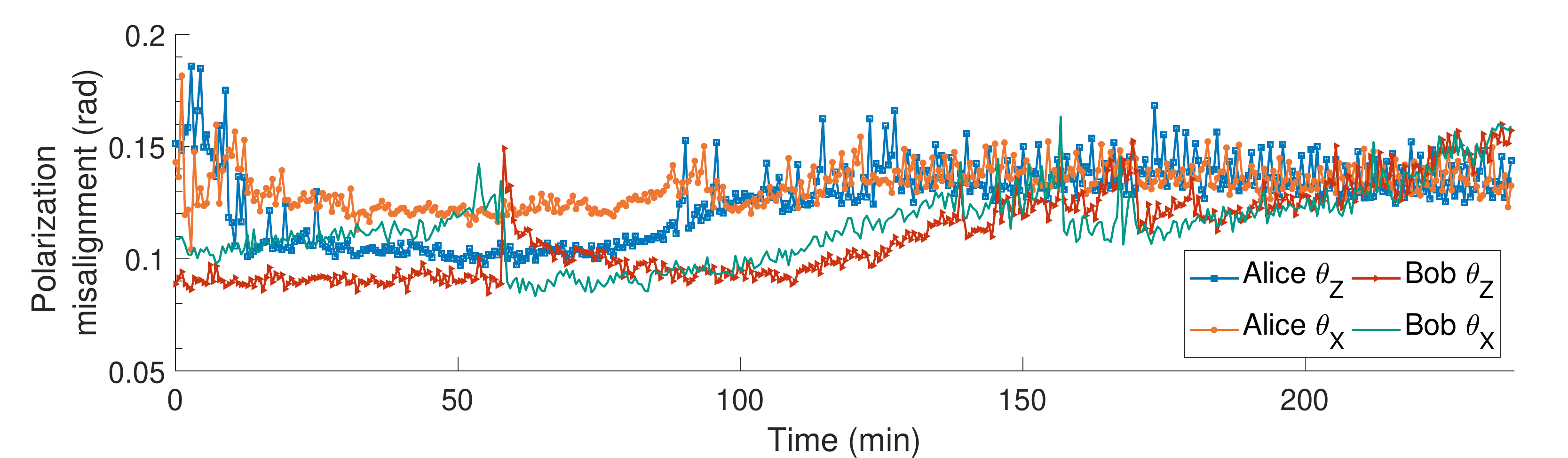}
\caption{}
  \label{track3}
\end{subfigure}%
\caption{Real-time tracking of polarization misalignment of the two users for an MDI-QKD session over 40 km of unisolated fibre with polarization compensation (a) Using optimized decoy intensities $\mu=0.28$, $\nu=0.07$, and $\omega=0.001$ with the respective probabilities $P_{\mu}=0.52$, $P_{\nu}=0.33$, and $P_{\omega}=0.15$. The average misalignment is maintained below 0.13 rad. (b) Using decoy settings $\mu=0.28,$ $\nu=0.07,$ $\omega=0.001,$ and $P_{\mu}=P_{\nu}=P_{\omega}=1/3$. The average misalignment is maintained below 0.13 rad. The average QBER is 3.8$\%$, and we achieved an average key rate of $7.45\times 10^{-6}$ bits per pulse.}
\label{fig:test}
\end{figure*}

We run the MDI-QKD experiment with real-time polarization compensation. We report experimental verification of our compensation scheme for two sets of decoy probabilities. 
In the first case the average photons per pulse for the signal and decoy intensities are $\mu=0.28$, $\nu=0.07$, and $\omega=0.001$ with the respective probabilities $P_{\mu}=0.52$, $P_{\nu}=0.33$, and $P_{\omega}=0.15$, where $P_{\mu}$ is the probability of sending a pulse with mean photon number $\mu$. These are numerically optimized probabilities based on the protocol presented in\cite{opt}. Our polarization compensation scheme maintained the average drift below 0.13 rad over
a period of three hours (Figure \ref{track2}). %Theoretically, this polarization misalignment could result in a QBER of 3.3 $\%$. 

Even though decoy settings were optimized and compensation ran successfully, we could not generate a key due to the limitation of one of our modules. Our arbitrary waveform generators can store a finite random pattern (1000 bits each for polarization and intensity selection), and the same pattern is repeated for continued operation. Any skew in the initial random pattern is amplified due to the many repetitions. Hence, the parameter optimization, which assumed continued randomness of users’ polarization and intensities for the entire duration of the experiment, is not ideal given our repetition of a short sequence of polarization and intensities. Note that the estimated number of coincidences corresponding to decoy intensities in the ideal case is already extremely small due to small decoy probabilities, intensities and the short experiment time. In addition to the low ideal decoy coincidence rate, the imperfect statistical fluctuations resulting from the limited size of the truly random sequence further prevented the key generation. Therefore, although the experiment using optimized decoy settings serves as a proof of concept for maintaining low polarization drift via our polarization compensation scheme, it is not ideal for key generation using our setup.

For this reason, we ran the experiment for modified parameters $\mu=0.28,$ $\nu=0.07,$ $\omega=0.001,$ and $P_{\mu}=P_{\nu}=P_{\omega}=1/3$ which has an increased value of decoy-state probabilities to obtain key generation as well as the same performance of polarization compensation. In the following, we report the data from a four hours MDI-QKD session with real-time polarization compensation with this latter choice of probabilities. Figure \ref{track3} shows the polarization misalignment in two bases for two users over four hours (estimated as $\arcsin(\text{error rate})$). The compensation scheme successfully maintained the average polarization misalignment below 0.13 rad for each user (Figure \ref{track3}). After the experiment, the two users perform sifting, error correction and privacy amplification to generate the secret key\footnote{Our analysis applies to imperfect channels with ideal BB84 states. To see the treatment of imperfect sources, see \cite{MDIQKD4,eli}.}. The secure key rate formula is given by \cite{MDIQKD, opt}

\begin{align}
R \geq p_{11}^Z Y_{11}^{Z,L}[1-H_2(e_{11}^{X,U})]-Q_{\mu \mu}^{Z}f(E_{\mu \mu}^Z)H_2(E_{\mu \mu}^{Z}),
\end{align}

where  $Q_{\mu \mu}^{Z}$ and $E_{\mu \mu}^{Z}$ are respectively the overall gain, and the error rate of signal states in the Z basis, $Y_{11}^{Z,L}$ is a lower bound on the yield of single-photon states in the  Z basis, and  $e_{11}^{X,U}$ is an upper bound on the phase error of single photon states. In our experiment, both $Y_{11}^{Z,L}$ and $e_{11}^{X,U}$ are estimated using an analytical method with two decoy states described in \cite{opt}. $f(E_{\mu \mu}^Z)=1.16$ is the efficiency of error correction and $H_2$ is the binary Shannon entropy function.

When using weak coherent pulses, since the QBER in the conjugate basis is high due to multiphoton pulses, only the projection basis can be used for key generation when selected by both Alice and Bob. Since we alternate the polarization beam splitter's axis between X and Z, we will have to extract the key separately from each basis. The successful Bell state measurements in the Z basis are compiled together to calculate the lower bound on the yield of single-photon states ($Y_{11}^{Z, L}$), the upper bound on the phase error rate of single-photon components ($e_{11}^{X, U}$), and the key rate $R_Z$ (Table \ref{ResultsZ}); similarly, the successful measurements in X basis are compiled together to calculate $Y_{11}^{X, L}$ and $e_{11}^{Z, U}$, and the key rate $R_X$ (Table \ref{ResultsX}). We were able to generate asymptotic key rates of $R_Z=5.94\times 10^{-6}$ and $R_X=8.96\times 10^{-6}$ bits per pulse.

In summary, we propose and implement a novel polarization compensation scheme for MDI-QKD by recycling some discarded detections. Our scheme evaluates the polarization drift in real-time based on single measurements corresponding to decoy intensities. The actuation solely relies on polarization controllers within the channel and the measurement setup required for alignment in MDI-QKD regardless of the implementation. The average polarization drift was successfully maintained below 0.13 rad over 40 km of spooled fibres left exposed and without isolation throughout a four-hour run. We achieved an average key rate of $7.45\times 10^{-6}$ bits per pulse.

Our current work on real-time polarization compensation paves the way for a polarization encoding MDI-QKD network since it can be implemented in real-time without reducing the key sharing cycle and retains a simple structure for a user node without requiring additional equipment. We also note that this scheme with minor modification is applicable for BB84 as the projections due to non-empty decoy states can be used to estimate the polarization drift and perform active compensation.

We thank Shihan Sajeed for insightful discussions. We also thank funding from NSERC, CFI, ORF, MITACS, US ONR, Royal Bank of Canada and Huawei Technologies Canada Inc., and the University of Hong Kong start-up grant.

\clearpage
\begin{table}[h]
\footnotesize
\begin{subtable}[b]{0.45\textwidth}
\centering
\begin{tabular}{|c | c c c |}   
\hline %$Gain_Z$ 
\diagbox{$I_A$}{$I_B$} & $\mu$ & $\nu$ & $\omega$ \\ [1ex]     
%heading
\hline    
$\mu$ & \makecell{$3.00\pm0.03$ \\ $\times 10^{-5}$} & \makecell{$8.06\pm0.13$\\ $\times 10^{-6}$} & \makecell{$7.56\pm0.41$ \\ $\times 10^{-7}$}\\  [1ex]
$\nu$ & \makecell{$9.23\pm0.13$\\ $\times 10^{-6}$}  & \makecell{$1.76\pm0.06$ \\ $\times 10^{-6}$}& \makecell{$4.39\pm1.01$ \\ $\times 10^{-8}$} \\  [1ex]
$\omega$ & \makecell{$5.62\pm0.35$\\ $\times 10^{-7}$}  & \makecell{$3.12\pm0.83$ \\ $\times 10^{-8}$} & 0 \\  [1ex]
\hline    
\end{tabular}
\caption{Experimental values of gains in Z basis}\label{GainZ}
\end{subtable}
\begin{subtable}[b]{0.45\textwidth}
\centering
\begin{tabular}{|c | c c c |}   
\hline %$Gain_Z$ 
\diagbox{$I_A$}{$I_B$} & $\mu$ & $\nu$ & $\omega$ \\ [1ex]     
%heading
\hline    
$\mu$ & \makecell{$6.11\pm0.04$ \\ $\times 10^{-5}$} & \makecell{$2.30\pm0.02$\\ $\times 10^{-5}$} & \makecell{$1.67\pm0.02$ \\ $\times 10^{-5}$}\\  [1ex]
$\nu$ & \makecell{$2.91\pm0.03$\\ $\times 10^{-5}$}  & \makecell{$3.33\pm0.09$ \\ $\times 10^{-6}$}& \makecell{$6.27\pm0.37$ \\ $\times 10^{-7}$} \\  [1ex]
$\omega$ & \makecell{$1.57\pm0.02$\\ $\times 10^{-5}$}  & \makecell{$6.85\pm0.39$ \\ $\times 10^{-7}$} & 0 \\  [1ex]
\hline    
\end{tabular}
\caption{Experimental values of gains in X basis}\label{GainXinZ}
\end{subtable}
\begin{subtable}[b]{0.45\textwidth}
\centering
\begin{tabular}{|c | c c c |} 
\hline 
\diagbox{$I_A$}{$I_B$} & $\mu$ & $\nu$ & $\omega$ \\ [1ex]     
%heading
\hline    
$\mu$ & 0.038$\pm$0.002  & 0.060$\pm$0.004 & 0.525$\pm$0.027 \\  [1ex]
$\nu$ & 0.056$\pm$0.004  & 0.043$\pm$0.007 & 0.458$\pm$0.114  \\  [1ex]
$\omega$ & 0.441$\pm$0.031  & 0.286$\pm$0.121 & N.A.  \\  [1ex]
\hline    
\end{tabular}
\caption{Experimental values of QBERs in Z basis}\label{QBERZ}
\end{subtable}
\begin{subtable}[b]{0.45\textwidth}
\centering
\begin{tabular}{|c | c c c |}   
\hline 
\diagbox{$I_A$}{$I_B$} & $\mu$ & $\nu$ & $\omega$ \\ [1ex]     
%heading
\hline    
$\mu$ & 0.279$\pm$0.003  & 0.355$\pm$0.005 & 0.489$\pm$0.006 \\  [1ex]
$\nu$ & 0.383$\pm$0.004  & 0.281$\pm$0.012 & 0.520$\pm$0.030  \\  [1ex]
$\omega$ & 0.495$\pm$0.006  & 0.472$\pm$0.029 & N.A.  \\  [1ex]
\hline    
\end{tabular}
\caption{Experimental values of QBERs in X basis}\label{PhaseZ}
\end{subtable}

\begin{subtable}[b]{0.45\textwidth}
\centering
\begin{tabular}{| c | c | c | c | c | c |} 
\hline 
\multicolumn{2}{|c|}{}&&&\multicolumn{2}{c|}{}\\
\multicolumn{2}{|c|}{ ${Y}^{11,L} (\times 10^{-4})$}&${e_x}^{11,U}$ &$R_\infty$ (bit&\multicolumn{2}{c|}{ $\theta_{avg}$ (rad)}\\
\cline{5-6}
\cline{1-2}
z & x  &  &  per pulse) & Alice&Bob\\ [1ex]
\hline    
8.02   &  9.91  & 0.148  & 5.94$\times 10^{-6}$ & 0.126 & 0.110  \\  [1ex]
\hline    
\end{tabular}
\caption{Estimated single photon yields, error rate, asymptotic key rate and the average polarization misalignments}\label{SubResultsZ}
\end{subtable}
\caption{Experimental results corresponding to the half of the experiment where the Bell state measurement includes projection unto Z basis.} \label{ResultsZ}
\end{table}

\begin{table}[h]
\footnotesize
\begin{subtable}[b]{0.45\textwidth}
\centering
\begin{tabular}{|c | c c c |}   
\hline %$Gain_Z$ 
\diagbox{$I_A$}{$I_B$} & $\mu$ & $\nu$ & $\omega$ \\ [1ex]     
\hline    
$\mu$ & \makecell{$3.19\pm0.03$ \\ $\times 10^{-5}$} & \makecell{$7.90\pm0.13$\\ $\times 10^{-6}$} & \makecell{$7.49\pm0.41$ \\ $\times 10^{-7}$}\\  [1ex]
$\nu$ & \makecell{$8.99\pm0.14$\\ $\times 10^{-6}$}  & \makecell{$1.81\pm0.06$ \\ $\times 10^{-6}$}& \makecell{$4.23\pm0.97$ \\ $\times 10^{-8}$} \\  [1ex]
$\omega$ & \makecell{$6.58\pm0.38$\\ $\times 10^{-7}$}  & \makecell{$4.68\pm1.02$ \\ $\times 10^{-8}$} & 0 \\  [1ex]
\hline    
\end{tabular}
\caption{Experimental values of gains in X basis}\label{GainX}
\end{subtable}
\begin{subtable}[b]{0.45\textwidth}
\centering
\begin{tabular}{|c | c c c |}   
\hline 
\diagbox{$I_A$}{$I_B$} & $\mu$ & $\nu$ & $\omega$ \\ [1ex]     
\hline    
$\mu$ & \makecell{$6.27\pm0.04$ \\ $\times 10^{-5}$} & \makecell{$2.20\pm0.02$\\ $\times 10^{-5}$} & \makecell{$1.66\pm0.02$ \\ $\times 10^{-5}$}\\  [1ex]
$\nu$ & \makecell{$2.99\pm0.03$\\ $\times 10^{-5}$}  & \makecell{$3.46\pm0.09$ \\ $\times 10^{-6}$}& \makecell{$7.29\pm0.41$ \\ $\times 10^{-7}$} \\  [1ex]
$\omega$ & \makecell{$1.58\pm0.02$\\ $\times 10^{-5}$}  & \makecell{$7.16\pm0.40$ \\ $\times 10^{-7}$} & 0 \\  [1ex]
\hline    
\end{tabular}
\caption{Experimental values of gains in Z basis}\label{GainZinX}
\end{subtable}
\begin{subtable}[b]{0.45\textwidth}
\centering
\begin{tabular}{|c | c c c |}   
\hline 
\diagbox{$I_A$}{$I_B$} & $\mu$ & $\nu$ & $\omega$ \\ [1ex]     
\hline    
$\mu$ & 0.038$\pm$0.002  & 0.069$\pm$0.004  & 0.509$\pm$0.027  \\  [1ex]
$\nu$ & 0.066$\pm$0.004   & 0.041$\pm$0.007  & 0.579$\pm$0.113   \\  [1ex]
$\omega$ & 0.475$\pm$0.029   & 0.524$\pm$0.109  & N.A.  \\  [1ex]
\hline    
\end{tabular}
\caption{Experimental values of QBERs in X basis}\label{QBERX}
\end{subtable}
\begin{subtable}[b]{0.45\textwidth}
\centering
\begin{tabular}{|c | c c c |}   
\hline 
\diagbox{$I_A$}{$I_B$} & $\mu$ & $\nu$ & $\omega$ \\ [1ex]     
\hline    
$\mu$ & 0.338$\pm$0.003  & 0.387$\pm$0.005 & 0.480$\pm$0.006 \\  [1ex]
$\nu$ & 0.382$\pm$0.004  & 0.269$\pm$0.011 & 0.523$\pm$0.028  \\  [1ex]
$\omega$ & 0.505$\pm$0.006  & 0.464$\pm$0.028 & N.A.  \\  [1ex]
\hline    
\end{tabular}
\caption{Experimental values of QBERs in Z basis}\label{PhaseX}
\end{subtable}
\begin{subtable}[b]{0.45\textwidth}
\centering
\begin{tabular}{| c | c | c | c | c | c |} 
\hline 
\multicolumn{2}{|c|}{}&&&\multicolumn{2}{c|}{}\\
\multicolumn{2}{|c|}{ ${Y}^{11,L} (\times 10^{-4})$}&${e_x}^{11,U}$ &$R_\infty$ (bit&\multicolumn{2}{c|}{ $\theta_{avg}$ (rad)}\\
\cline{5-6}
\cline{1-2}
z & x  &  &  per pulse) & Alice&Bob\\ [1ex]
\hline   
9.82 &  8.17   & 0.117   & 8.96$\times 10^{-6}$ & 0.132  & 0.115 \\  [1ex]
\hline 
\end{tabular}
\caption{Estimated single photon yields, error rate, asymptotic key rate and the average polarization misalignments}\label{SubResultsX}
\end{subtable}
\caption{Experimental results corresponding to the half of the experiment where the Bell state measurement includes projection unto X basis.} \label{ResultsX}
\end{table}


\clearpage

\begin{thebibliography}{50}
\small

\bibitem{Security}
Bennett, C. H., Brassard, G. "Quantum cryptography: Public key distribution and coin tossing."
\href{https://doi.org/10.48550/arXiv.2003.06557}{ Proc. Int. Conf. on Computers, Systems Signal Process, pp. 175-179 (1984)}; Ekert, A. K. "Quantum cryptography based on Bell's theorem." \href{https://doi.org/10.1103/PhysRevLett.67.661}{ Phys. Rev. Lett. 67, 661 (1991)}

\bibitem{hacking}
Xu, Feihu, et al. "Secure quantum key distribution with realistic devices." 
\href{https://doi.org/10.1103/RevModPhys.92.025002}{Rev. Mod. Phys. 92.2 (2020): 025002}; Pirandola, Stefano, et al. "Advances in quantum cryptography."
\href{https://doi.org/10.1364/AOP.361502}{Adv. Opt. Photonics 12.4 (2020): 1012-1236.}

\bibitem{MDIQKD}
Lo, H.-K., et al. "Measurement-device-independent quantum key distribution."
\href{https://doi.org/10.1103/PhysRevLett.108.130503}{Phys. Rev. Lett. 108, 130503  (2012).}

\bibitem{MDIQKD4}
Tang, Z., et al. "Experimental demonstration of polarization encoding measurement-device-independent quantum key distribution." \href{https://doi.org/10.1103/PhysRevLett.112.190503}{Phys. Rev. Lett. 112.19 (2014): 190503}. 

\bibitem{MDIQKD3}
Tang, Z., et al. "Experimental measurement-device-independent quantum key distribution with imperfect sources." \href{https://doi.org/10.1103/PhysRevA.93.042308}{Phys. Lett. A 93.4 (2016): 042308}. 

\bibitem{MDIQKD1}
Liu, Yang, et al. "Experimental measurement-device-independent quantum key distribution." \href{https://doi.org/10.1103/PhysRevLett.111.130502}{Phys. Rev. Lett. 111.13 (2013): 130502}.

\bibitem{MDIQKD2}
Tang, Yan-Lin, et al. "Measurement-device-independent quantum key distribution over 200 km." \href{https://doi.org/10.1103/PhysRevLett.113.190501}{Phys. Rev. Lett. 113.19 (2014): 190501}.

\bibitem{MDIQKD5}
Tang, Yan-Lin, et al. "Measurement-device-independent quantum key distribution over untrustful metropolitan network." \href{https://doi.org/10.1103/PhysRevX.6.011024}{Phys. Rev. X 6.1 (2016): 011024}. 

\bibitem{MDIQKD6}
Tang, Yan-Lin, et al. "Field test of measurement-device-independent quantum key distribution." \href{https://doi.org/10.1109/JSTQE.2014.2361796}{IEEE J. Sel. Top. Quantum Electron. 21.3 (2014): 116-122}. 

\bibitem{MDIQKD7}
Liu, Hui, et al. "Experimental demonstration of high-rate measurement-device-independent quantum key distribution over asymmetric channels." \href{https://doi.org/10.1103/PhysRevLett.122.160501}{Phys. Rev. Lett. 122.16 (2019): 160501}.
 
\bibitem{MDIQKD8}
Wang, Chao, et al. "Measurement-device-independent quantum key distribution robust against environmental disturbances." \href{https://doi.org/10.1364/OPTICA.4.001016}{Optica 4.9 (2017): 1016-1023}. 

\bibitem{MDIQKD9}
Wang, Chao, et al. "Phase-reference-free experiment of measurement-device-independent quantum key distribution." \href{https://doi.org/10.1103/PhysRevLett.115.160502}{Phys. Rev. Lett. 115.16 (2015): 160502}. 

\bibitem{MDIQKD10}
Rubenok, Allison, et al. "Real-world two-photon interference and proof-of-principle quantum key distribution immune to detector attacks." \href{https://doi.org/10.1103/PhysRevLett.111.130501}{Phys. Rev. Lett. 111.13 (2013): 130501}.

\bibitem{MDIQKD11}
Da Silva, T. Ferreira, et al. "Proof-of-principle demonstration of measurement-device-independent quantum key distribution using polarization qubits." \href{https://doi.org/10.1103/PhysRevA.88.052303}{Phys. Rev. A 88.5 (2013): 052303}.

\bibitem{MDIQKD12}
Valivarthi, Raju, et al. "Measurement-device-independent quantum key distribution: from idea towards application." \href{https://doi.org/10.1080/09500340.2015.1021725}{J. Mod. Opt. 62.14 (2015): 1141-1150}.

\bibitem{MDIQKD13}
Pirandola, Stefano, et al. "High-rate measurement-device-independent quantum cryptography." \href{https://doi.org/10.1038/nphoton.2015.83}{Nat. Photonics 9.6 (2015): 397-402}.

\bibitem{MDIQKD14}
Yin, Hua-Lei, et al. "Measurement-device-independent quantum key distribution over a 404 km optical fiber." \href{https://doi.org/10.1103/PhysRevLett.117.190501}{Phys. Rev. Lett. 117.19 (2016): 190501}.

\bibitem{MDIQKD15}
Comandar, L.C., et al. "Quantum key distribution without detector vulnerabilities using optically seeded lasers." \href{https://doi.org/10.1038/nphoton.2016.50}{Nat. Photonics 10.5 (2016): 312}.

\bibitem{MDIQKD16}
Kaneda, Fumihiro, et al. "Quantum-memory-assisted multiphoton generation for efficient quantum information processing." \href{https://doi.org/10.1364/OPTICA.4.001034}{ Optica 4.9 (2017): 1034-1037}.

\bibitem{MDIQKD17}
Valivarthi, Raju, et al. "A cost-effective measurement-device-independent quantum key distribution system for quantum networks." \href{https://doi.org/10.1088/2058-9565/aa8790}{Quantum Sci. Technol.  2.4 (2017): 04LT01}.

\bibitem{MDIQKD18}
Liu, Hongwei, et al. "Polarization-multiplexing-based measurement-device-independent quantum key distribution without phase reference calibration." \href{https://doi.org/10.1364/OPTICA.5.000902}{Optica 5.8 (2018): 902-909}.

\bibitem{MDIQKD19}
Wei, Kejin, et al. "High-Speed Measurement-Device-Independent Quantum Key Distribution with Integrated Silicon Photonics." \href{https://doi.org/10.1103/PhysRevX.10.031030}{Phys. Rev. X 10, 031030  (2020)}.

\bibitem{MDIQKD20}
Yuan, Yi-ping, et al. "Proof-of-principle demonstration of measurement-device-independent quantum key distribution based on intrinsically stable polarization-modulated units." \href{https://doi.org/10.1364/OE.387968}{Opt. Express 28.8 (2020): 10772-10782.}.

\bibitem{MDIQKD21}
Zhou, Xing-Yu, et al. "Experimental three-state measurement-device-independent quantum key distribution with uncharacterized sources." \href{https://doi.org/10.1364/OL.398993}{Opt. Lett. Vol. 45, Issue 15, pp. 4176-4179 (2020)}.

\bibitem{Drift}
Y.Y. Ding, et al., "Polarization variations in installed fibers and their influence on quantum key distribution systems" 
\href{https://doi.org/10.1364/OE.25.027923}{Opt. Express 25, 27923-27936 (2017)}

\bibitem{PolComp4}
Ma, Lijun, Hai Xu, and Xiao Tang. "Polarization recovery and auto-compensation in quantum key distribution network." \href{https://doi.org/10.1117/12.679575}{Proc. SPIE 6305, 630513 (2006)}.

\bibitem{PolComp5}
Tang, Xiao, et al. "Experimental study of high speed polarization-coding quantum key distribution with sifted-key rates over Mbit/s." \href{https://doi.org/10.1364/OE.14.002062}{Opt. Express 14.6 (2006): 2062-2070}.

\bibitem{PolComp6}
Chen, Jie, et al. "Active polarization stabilization in optical fibers suitable for quantum key distribution." \href{https://doi.org/10.1364/OE.15.017928}{Opt. Express 15.26 (2007): 17928-17936}.

\bibitem{PolComp9}
Peng, Cheng-Zhi, et al. "Experimental long-distance decoy-state quantum key distribution based on polarization encoding." 
\href{https://doi.org/10.1103/PhysRevLett.98.010505}{Phys. Rev. Lett. 98.1 (2007): 010505.}.

\bibitem{PolComp24}
Wu, Guang, et al. "Stable polarization-encoded quantum key distribution in fiber." 
\href{https://arxiv.org/abs/quant-ph/0606108 }{arXiv preprint quant-ph/0606108 (2006) }.

\bibitem{PolComp13}
Ma, Lijun, Alan Mink, and Xiao Tang. "High speed quantum key distribution over optical fiber network system." 
\href{https://doi.org/10.6028/jres.114.010}{J. Res. Natl. Inst. Stand. Technol. 114.3 (2009): 149}.

\bibitem{PolComp12}
Liu, Yang, et al. "Decoy-state quantum key distribution with polarized photons over 200 km." 
\href{https://doi.org/10.1364/OE.18.008587}{Opt. Express 18.8 (2010): 8587-8594}.

\bibitem{PolComp10}
Xavier, G. B., et al. "Experimental polarization encoded quantum key distribution over optical fibres with real-time continuous birefringence compensation." \href{https://doi.org/10.1088/1367-2630/11/4/045015}{New J. Phys. 11.4 (2009): 045015}.

\bibitem{PolComp19}
Yoshino, Ken-ichiro, et al. "Maintenance-free operation of WDM quantum key distribution system through a field fiber over 30 days." 
\href{https://doi.org/10.1364/OE.21.031395 }{Opt. Express 21.25 (2013): 31395-31401}.

\bibitem{PolComp22}
Xavier, Guilherme Barreto, et al. "Active polarization control for quantum communication in long‐distance optical fibers with shared telecom traffic." 
\href{ https://doi.org/10.1002/mop.26320}{Microw. Opt. Technol. Lett. 53.11 (2011): 2661-2665 }.

\bibitem{PolComp23}
Li, Dong-Dong, et al. "Field implementation of long-distance quantum key distribution over aerial fiber with fast polarization feedback."
\href{https://doi.org/10.1364/OE.26.022793 }{Opt. Express 26.18 (2018): 22793-22800 }.

\bibitem{PolComp18}
Muga, Nelson J., et al. "Optimization of polarization control schemes for QKD systems." 
\href{https://doi.org/10.1117/12.892221 }{Proc. SPIE 8001, 80013N (2011)}.

\bibitem{PolComp3}
Yuan, Z. L., and A. J. Shields. "Continuous operation of a one-way quantum key distribution system over installed telecom fibre." \href{https://doi.org/10.1364/OPEX.13.000660}{Opt. Express 13.2 (2005): 660-665}.

\bibitem{PolComp11}
Chen, J., et al. "Stable quantum key distribution with active polarization control based on time-division multiplexing."
\href{https://doi.org/10.1088/1367-2630/11/6/065004}{New J. Phys. 11.6 (2009): 065004}.

\bibitem{PolComp21}
Yu, Shengrong Timothy, et al. "Adaptive polarization-state monitoring and stabilization scheme for one-way polarization-encoded quantum key distribution systems."
\href{https://doi.org/10.1109/CLEOPR.2015.7376021 }{2015 Conf. Lasers Electro-Opt. Pac. Rim CLEO-PR, 2015, pp. 1-2}.

\bibitem{PolComp16}
Pugh, Christopher J., et al. "Airborne demonstration of a quantum key distribution receiver payload." 
\href{ https://doi.org/10.1088/2058-9565/aa701f}{Quantum Sci. Technol. 2.2 (2017): 024009}.

\bibitem{PolComp20}
Almeida, Álvaro J., et al. "Continuous control of random polarization rotations for quantum communications." 
\href{https://doi.org/10.1109/JLT.2016.2587818 }{J. Light. Technol. 34.16 (2016): 3914-3922}.

\bibitem{PolComp25}
Agnesi, Costantino, et al. "Simple Quantum Key Distribution with qubit-based synchronization and a self-compensating polarization encoder."
\href{https://doi.org/10.1364/OPTICA.381013}{Optica 7.4 (2020): 284-290}.

\bibitem{PolComp17}
Ding, Yu-Yang, et al. "Polarization-basis tracking scheme for quantum key distribution using revealed sifted key bits." 
\href{https://doi.org/10.1364/OL.42.001023 }{Opt. Lett. 42.6 (2017): 1023-1026}.

\bibitem{discarded}
Barnett, Stephen M., and Simon JD Phoenix. "Bell's inequality and rejected-data protocols for quantum cryptography." \href{https://doi.org/10.1080/09500349314551511}{J. Mod. Opt. 40.8 (1993): 1443-1448}.

\bibitem{thesis}
Bedroya, O., 2023. Resource-Efficient Real-Time Polarization Compensation for MDI-QKD (Doctoral dissertation, University of Toronto (Canada)).
\bibitem{decoy}
Hwang, Won-Young. "Quantum key distribution with high loss: toward global secure communication." 
\href{https://doi.org/10.1103/PhysRevLett.91.057901}{Phys. Rev. Lett. 91.5 (2003): 057901}. 
 
Lo, Hoi-Kwong, et al. "Decoy state quantum key distribution." 
\href{https://doi.org/10.1103/PhysRevLett.94.230504}{Phys. Rev. Lett. 94.23 (2005): 230504}. 
 
Wang, Xiang-Bin. "Beating the photon-number-splitting attack in practical quantum cryptography."
\href{https://doi.org/10.1103/PhysRevLett.94.230503}{Phys. Rev. Lett. 94.23 (2005): 230503}. 
 
\bibitem{PolM}
Lucio-Martinez, I., et al. "Proof-of-concept of real-world quantum key distribution with quantum frames." \href{https://doi.org/10.1088/1367-2630/11/9/095001}{New J. Phys. 11.9 (2009): 095001}. 
 
\bibitem{opt}
Xu, Feihu, et al. "Practical aspects of measurement-device-independent quantum key distribution." 
\href{https://doi.org/10.1088/1367-2630/15/11/113007}{New J. Phys. 15.11 (2013): 113007};

\bibitem{coiled}
R. Ulrich, et al. "Bending-induced birefringence in single-mode fibers." 
\href{https://doi.org/10.1364/OL.5.000273}{Opt. Lett. 5, 273-275 (1980)}
\bibitem{coiled2}
S. C. Rashleigh and R. Ulrich, "High birefringence in tension-coiled single-mode fibers." 
\href{https://doi.org/10.1364/OL.5.000354}{Opt. Lett. 5, 354-356 (1980)}


\bibitem{Bell}
K. Mattle et al., "Dense Coding in Experimental Quantum Communication" 
\href{https://doi.org/10.1103/PhysRevLett.76.4656}{Phys. Rev. Lett. 76, 4656 (1996)}

\bibitem{chernoff2}
Curty, Marcos, et al. "Finite-key analysis for measurement-device-independent quantum key distribution." \href{https://doi.org/10.1038/ncomms4732}{Nat. Commun. 5.1 (2014): 3732.}



\bibitem{Chernoff}
 Motwani, Rajeev, and Prabhakar Raghavan. Randomized algorithms. \href{https://doi.org/10.1017/CBO9780511814075}{Cambridge university press, 1995.}

 \bibitem{eli}
 Bourassa, J. Eli, et al. "Loss-tolerant quantum key distribution with mixed signal states." \href{https://doi.org/10.1103/PhysRevA.102.062607}{Phys. Rev. A 102.6 (2020): 062607.}
 
\end{thebibliography}
\end{document}